\documentstyle[12pt]{article}
\begin{document}

\centerline{\bf Gauge Parameter Dependence in
Gauge Theories}

\bigskip

\centerline{J. Gegelia, A. Khelashvili and N. Kiknadze}

\medskip

\centerline{\it High Energy Physics
Institute, University str.\ 9, 380086 Tbilisi, Republic of Georgia}

\medskip

\centerline{\bf Abstract}
\medskip

\begin{abstract}
On the example of topologically massive gauge field theory we find
the origin of possible inconsistency of working with gauge fixing terms
(together with relevant ghost sector).
\end{abstract}

\bigskip

The common procedure of practical calculations in e.g.\ QED and QCD
is the introduction of gauge-fixing terms into Lagrangian together
with the relevant ghost sector. It is understood that these terms
should not influence physical quantities. However, encountering gauge
parameter dependence of some quantity it is not easy to understand
whether it is the gauge fixing procedure that is inconsistent or it
is the fault of the theory itself.
Below we are going to show
that possible source of inconsistency of working with gauge fixing
terms resides in averaging the generating functional for Green's
functions for different gauges. Accordingly, when the mentioned
procedure of averaging is inconsistent, the
quantization in Hamilton formalism is not equivalent to
the Lagrange formalism quantization (with the help of introduction
of gauge fixing terms).

The quantization rules for gauge theories in Lagrange formalism were
derived in \cite{1}. It is well known that this rules must be modified
for more general theories \cite{2}. The main method to construct a
covariant gauge quantum theory is the BRST quantization \cite{3}.
Equivalence of the Lagrangian and Hamiltonian approaches for the
BRST quantization was demonstrated in \cite{4}.

For an illustration we are going to use an example where the Faddev-Popov
quantization in the Hamilton formalism produces correct results, while
quantization in the Lagrange formalism by adding covariant gauge
fixing term leads to the gauge parameter dependence of physical
quantities.

Consider topologically massive abelian gauge field coupled to
fermion (in 2+1 dimensions) \cite{5}:
\begin{equation}
L=-\frac{1}{4}F_{\mu\nu}F^{\mu\nu}+\frac{M}{4}\epsilon_{\mu\nu\lambda}
A^{\mu}F^{\nu\lambda}+\bar\psi(i\hat\partial+
e\hat A-m)\psi .                     \label{1}
\end{equation} %1
In \cite{5} it was noticed that the fermion pole mass calculated in
the covariant gauge (i.e.\ $L_{gf}=(1/2\xi)(\partial A)^2$) at one
loop depends on the gauge fixing parameter $\xi$.
The wave function renormalization constant as well as the
{\cal S}-matrix elements are infrared (IR) divergent in any covariant
gauge other then the Landau one.
Authors of \cite{5} found problematic to define the
mass--shell in this theory in gauge invariant way. We are
going to show that in fact these gauge parameter dependence problems
are due to the inconsistency of the gauge fixing procedure.

We have checked by explicit calculations that the fermion pole mass in
the Coulomb and axial gauges equals to the Landau gauge ($\xi=0$) one.
(In \cite{5} it was noted that the IR safe version of the
gauge field propagator in the Coulomb gauge produces the Landau
gauge pole mass. We just checked that the same is true for the full
propagator.) Besides, the {\cal S}-matrix elements, being finite in the
Landau (and also
in the Coulomb and axial gauges), suffer from severe IR divergences in any
other covariant ($\xi\neq 0$) gauge.

To investigate the origin of this apparent puzzle we have performed
the canonical quantization of this theory in the physical (gauge
invariant) variables and found that the resulting quantum theory
coincides with canonically quantized theory in Coulomb gauge. We do
not reproduce quantization in physical variables here --- it is
analogous to the ordinary QED case \cite{gitman}.

So, for this theory, the Coulomb gauge is directly related to the
quantization in gauge invariant variables and hence any other
consistent quantization scheme must reproduce the Coulomb gauge results.

In the Lorentz invariant gauge $\partial_\mu A^\mu=C(x)$ the generating
functional for the Green's functions is given in the form \cite{cabo}:
\begin{equation}
Z_C[J,\zeta,\bar\zeta]=\int{\cal D}A_\mu{\cal D}\psi{\cal D}\bar\psi
\delta\left(\partial_\mu A^\mu-C(x)\right) det{\cal M}_L
e^{iS+i\int dx (J^\mu A_\mu+\bar\psi\zeta+\bar\zeta\psi)} .
\label{2}
\end{equation}                                  %2
It is easy to see that for the choice $C(x)=0$ the Green's functions
generated by (\ref{2}) are identical to the Landau gauge ones. So both
Landau and Coulomb gauges are implied from Hamilton formalism quantization.
Note that in our example problems start in the covariant non--Landau
gauges.

Now let us recall how the Lagrange formalism quantization by adding
the gauge fixing terms to the Lagrangian can be reproduced from (\ref{2}).
Multiply (\ref{2}) by $exp\{\frac{i}{2\xi}\int C^2(x)dx\}$ and integrate
over $C(x)$. We get:
\begin{equation}
Z[J,\zeta,\bar\zeta]=\int{\cal D}A_\mu{\cal D}\psi{\cal D}\bar\psi
det{\cal M}_L
e^{i\int dx \left\{L
+
\frac{1}{2\xi}(\partial A)^2+J^\mu A_\mu+\bar\psi\zeta+\bar\zeta\psi)
\right\}} .        \label{3}
\end{equation}
Eq.~(\ref{3}) is an expression for the generating functional which is
usually used in the framework of the Lagrange quantization. We gave this
formal
derivation of (\ref{3}) because it clearly indicates that the addition of
gauge fixing terms amounts to the averaging of different gauges
$\partial_\mu A^\mu=C(x)$ with the weight $exp\{\frac{i}{2\xi}\int
C^2(x)dx\}$. In fact it does not correspond to any definite gauge at
all \cite{sl}, rather it is a mixture of different gauges.

For any definite $C(x)$, $Z_C$ leads to the $C(x)$-independent
physical quantities. Usually it is conjectured that averaging $Z_C$
for different $C(x)$ will translate this $C(x)$ independence into
the gauge parameter --- $\xi$ independence. We claim that this
averaging procedure may ruin
equivalence of (\ref{2}) and (\ref{3}).  The situation is similar
to the infinite sum of vanishing terms --- the sum may turn out to
be finite or even divergent.

Indeed let us examine the $C(x)$ dependent gauge field propagator in
the model (\ref{1}):
\begin{equation}
D^C_{\mu\nu}=\frac{1}{N}\int{\cal D}A_\mu{\cal D}\psi{\cal D}\bar\psi
\delta\left(\partial_\mu A^\mu-C(x)\right) A_\mu A_\nu
e^{iS} .
\label{4}
\end{equation}
Let us perform gauge transformation $A_\mu\to A_\mu+\partial_\mu
\partial^{-2}C(x)$:
$$
D^C_{\mu\nu}(p,q)=\delta(p-q)D^L_{\mu\nu}(p)+C(p)C(q)\frac{p_\mu
q_\nu}{p^2q^2} .
$$
Here $D^L$ denotes propagator in the Landau gauge and we have
passed to the momentum space. (The function $C(x)$ breaks translational
invariance and results in the nondiagonal term.)
Now, if we define the fermion mass and wave function renormalization
constant from the diagonal part, it is easy to see that no IR problems
will arise while calculating the physical amplitudes. (Although it seems
trivial from simple power counting, we explicitly checked it for
fermion scattering in the external field.)
If we integrate (\ref{4}) over $C$ with the weight
$exp\{\frac{i}{2\xi}\int C^2(x)dx\}$ we will get the usual
propagator corresponding to the gauge fixing Lagrangian
$(1/2\xi)(\partial_\mu A^\mu)^2$:
\begin{equation}
 D_{\mu\nu}=\frac{-i}{p^2-M^2}(g_{\mu\nu}-\frac{p_\mu p_\nu}{p^2}
-\frac{iM}{p^2}\epsilon_{\mu\nu\lambda}p^\lambda)+\xi\frac{p_\mu
p_\nu}{p^4} .
\end{equation}
Note the singular behaviour of the $\xi$-dependent part. It
causes gauge parameter dependence of the fermion pole mass and
also IR divergences of the ${\cal S}$-matrix elements.
Contribution of this term to the fermion pole mass is proportional
to:
\begin{equation}
\delta m_\xi\sim (p^2-m^2)^2\left.\int
d^nq\frac{1}{q^4\left((p-q)^2-m^2\right)} \right\vert_{p^2\to
m^2, n\to 3} .
\end{equation}
Due to the infrared divergence of the integral this expression will
yield finite gauge parameter dependent contribution to the mass. In
the same manner there arise IR divergences in the {\cal S}-matrix
elements, which are absent in the physical Coulomb, axial and Landau gauges.
Obviously in the described situation the LSZ formalism does not commute
with the integration over $C$. Indeed, if we calculate Green's functions
in $\partial_{\mu}A^{\mu}(x)=C(x)$ gauge and calculate physical
quantities using LSZ formalism, we find that they are well defined and do
not depend on $C(x)$. Integration over $C(x)$ leads only to the numerical
factor which in fact is cancelled by the normalization factor. On the
other hand, integrating $Z_C[J,\zeta ,\bar{\zeta}]$ over $C(x)$ and
applying LSZ formulas we find that physical quantities are IR divergent
and $\xi$-dependent. So, LSZ formalism and integration over $C(x)$ do not
commute in this particular case (and hence, in general).

The generating functional (\ref{3}) may be derived also with the help of
BRST quantization. So it provides us with one more example \cite{gov}
when the BRST quantization fails.

This kind of effect may happen in four dimensions too if the
averaging weight function is chosen in the form \cite{sl}:
\begin{equation}
exp\{\frac{i}{2\xi}\int [f(\partial^2)C]^2(x)dx\}
\label{wf}
\end{equation}
where $f$ is an arbitrary function of D'Alambert operator. Choosing
$f=(\partial^2)^k$ will lead to $(q^2)^{-(k+1)}$ singularity in the
gauge field propagator and hence result additional (incurable
without some artificial {\it ad hoc} procedure) infrared problems in
ordinary QED and QCD.

Of course one may try to foresee in what circumstances will the
described problem arise. Consider again the covariant gauge. In the
propagator of gauge particle the
tensor structure of the gauge fixing term ($p_\mu p_\nu$) is present
also in the Landau gauge propagator. Evidently problems begin if the
IR behaviour of the coefficient of this tensor structure in gauge fixing part
is more singular then the same in the Landau gauge. It is clear enough
because, normally, due to the gauge invariance, such structure will not
affect physical quantities. Sometimes comparison of IR behaviour of terms
of the bare propagator may be misleading. E.g., at the first sight,
the above criterion is not satisfied for conventional QED in three
dimensions. However, in this theory the Chern-Simons term is
generated dynamically \cite{5} (discussion of regularization
dependence of this fact noted in \cite{5}, can be found in
\cite{khel}) changing the IR behaviour of the Landau gauge terms.
Similar analysis must be applied to the other gauge fixing conditions
together with weight functions in order find when does the averaging
procedure introduce additional higher order IR singularities.

So we have demonstrated that at certain cases Lagrange and Hamilton
formalism quantizations are not equivalent due to inadmissibility
of averaging different gauge fixing conditions.
Hence tests of gauge parameter independence
of the physical quantities not always serve for checking consistency of
the theory, but rather represent a test for applicability of
Lagrange quantization with gauge fixing Lagrangian and of BRST
quantization. The possible source of inconsistency
of calculations with gauge fixing Lagrangeans resides in the
seemingly `innocent' procedure --- averaging different
gauges with some weight (which is equivalent to addition of the gauge
fixing terms).

Acknowledgements: we would like to thank G. Chechelashvili,
G. Jorjadze and B. Magradze for helpful discussions.

\end{document}